\documentclass[preprint,showpacs,preprintnumbers,amsmath,amssymb]{revtex4}

\usepackage{graphicx}% Include figure files
\usepackage{dcolumn}% Align table columns on decimal point
\usepackage{bm}% bold math

\begin{document}

\title{Strange Nucleon Form Factors from $ep$ and $\nu p$ Elastic Scattering}

\author{Stephen Pate}
\email{pate@nmsu.edu}
\affiliation{Physics Department, New Mexico State University, 
Las Cruces NM 88003, USA}

%\date{\today}%

\begin{abstract}
The recent parity-violating $ep$ forward-scattering elastic asymmetry
data from Jefferson Lab (HAPPEx and G0), when combined with the $\nu
p$ elastic cross section data from Brookhaven (E734), permit an
extraction of the strangeness contribution to the vector and axial
nucleon form factors for momentum transfers in the range $0.45 < Q^2 <
1.0$ GeV$^2$.  These results, combined with the recent determination
of the strange vector form factors at $Q^2 = 0.1$ GeV$^2$ (SAMPLE,
HAPPEx, PVA4, G0) have been interpreted in terms of $uuds\bar{s}$
configurations very different from the kaon-loop configurations
usually associated with strangeness in the nucleon.  New experiments
are being proposed to improve the state of our knowledge of the $\nu
p$ elastic cross section --- these new experiments will push the range
of $Q^2$ to much lower values, and greatly increase the precision of
the $\nu p$ elastic data.  One outcome of this can be a measurement of
the strangeness contribution to the nucleon spin, $\Delta s$.  Nuclear
targets (e.g. C or Ar) are to be used in these neutrino experiments,
and so a deep understanding of the nuclear physics, particularly in
regard to final state effects, is needed before the potential of these
precision experiments can be fully realized.
\end{abstract}


\maketitle

\section{Introduction}

Ever since the discovery of the first ``strange'' particles in cosmic
ray experiments~\cite{Rochester:1947mi} and the subsequent formulation
of the 3-quark model of baryons~\cite{Gell-Mann:1964nj}, nuclear and
particle physicists have sought to understand the role the strange
quark plays in ``non-strange'' particles like the proton.
Traditionally these investigations have taken place in the context of
deep-inelastic scattering. 
However, a strong effort has been made to measure the strange quark
contribution to the elastic form factors of the proton, in particular
the vector (electric and magnetic) form factors.  These
experiments~\cite{Mueller:1997mt,Hasty:2001ep,Spayde:2003nr,
Ito:2003mr,Aniol:2004hp,Maas:2004ta,Maas:2004dh,
Armstrong:2005hs,HAPPEx_1H_010,HAPPEx2006} exploit an interference
between the $\gamma$-exchange and $Z$-exchange amplitudes in order to
measure weak elastic form factors $G_E^{Z,p}$ and $G_M^{Z,p}$ which
are the weak-interaction analogs of the more traditional
electromagnetic elastic form factors $G_E^{\gamma,p}$ and
$G_M^{\gamma,p}$ for which copious experimental data are available.
The interference term is observable as a parity-violating asymmetry in
elastic $\vec{e}p$ scattering, with the electron longitudinally
polarized. By combining the electromagnetic form factors of the proton
and neutron with the weak form factors of the proton, one may separate
the up, down, and strange quark contributions; for example, the
electric form factors may be written as follows:
\begin{eqnarray*}
G_E^{\gamma,p} &=& \frac{2}{3}G_E^u - \frac{1}{3}G_E^d - \frac{1}{3}G_E^s \\
G_E^{\gamma,n} &=& \frac{2}{3}G_E^d - \frac{1}{3}G_E^u - \frac{1}{3}G_E^s \\
G_E^{Z,p} &=& \left(1-\frac{8}{3}\sin^2\theta_W\right)G_E^u
+\left(-1+\frac{4}{3}\sin^2\theta_W\right)G_E^d
+\left(-1+\frac{4}{3}\sin^2\theta_W\right)G_E^s.
\end{eqnarray*}
There is an assumption of charge symmetry, and also an assumption that the
role played by the strange quarks in the proton and neutron is the same.

Because of the weak-interaction process at the heart of this measurement
program, these parity-violating asymmetries also involve the axial form 
factor of the proton, which in a pure weak-interaction process takes
this form:
$$G_A^{Z,p} = \frac{1}{2}\left(-G_A^u + G_A^d + G_A^s\right).$$
The $u-d$ portion of this form factor is well-known from
neutron $\beta$-decay and other charged-current ($CC$) weak 
interaction processes like $\nu_\mu + n \rightarrow p+ \mu^-$:
$$G_A^{CC} = G_A^u-G_A^d = \frac{g_A}{(1+Q^2/M_A^2)^2}$$
where $g_A = 1.2695 \pm 0.0029$ is the axial coupling 
constant~\cite{PDG2004} and
$M_A = 1.001 \pm 0.020$ is the so-called ``axial mass'' which is
a very successful fitting parameter for the data 
on this form factor~\cite{Bodek:2003ed}.
The strange quark portion, $G_A^s$, is essentially unknown, except for some 
contradictory indications from polarized deep-inelastic scattering, 
which we discuss
in more detail below.  In elastic $ep$ scattering, the axial form 
factor does not
appear as a pure weak-interaction process; there are significant radiative
corrections which carry non-trivial theoretical uncertainties.  The result is
that, while the measurement of parity-violating asymmetries in 
$\vec{e}p$ elastic scattering is well suited to a measurement of 
$G_E^s$ and $G_M^s$, these experiments cannot cleanly extract $G_A^s$.  The
strange axial form factor is of great interest, however, because of the role
it plays in the understanding of the spin structure of the proton.

\section{Strange Quarks and the Spin of the Proton}

The strange quark contribution to the proton spin has been a subject of 
investigation ever since the first polarized inclusive deep-inelastic
measurements of the spin-dependent structure function $g_1(x)$ 
by EMC \cite{Ashman:1989ig}
demonstrated that the Ellis-Jaffe sum 
rule \cite{Ellis:1973kp,PhysRevD.10.1669.4} did not hold true.
Subsequent measurements at CERN and SLAC supported the initial EMC 
measurements, and a global analysis \cite{Filippone:2001ux} of these data
suggested $\Delta s \approx -0.15$.  This analysis carries with it an
unknown theoretical uncertainty because the deep-inelastic data must be
extrapolated to $x=0$ and an assumption of SU(3)-flavor 
symmetry must be invoked.

In the meantime, the E734 experiment \cite{Ahrens:1987xe} at Brookhaven 
measured the $\nu p$ and $\bar{\nu} p$ elastic scattering cross sections
in the momentum-transfer range $0.45<Q^2<1.05$ GeV$^2$.  These cross sections
are very sensitive to the strange axial form factor of the proton,
$G_A^s(Q^2)$, which is related to the strange quark contribution to the
proton spin:  $G_A^s(Q^2=0)=\Delta s$.  Assuming the strange axial form
factor had the same $Q^2$-dependence as the isovector axial form factor,
E734 also extracted a negative value for $\Delta s$.  However, this
determination was hampered by the large systematic uncertainies in the
cross section measurement, as well as a lack of knowledge of the strange
vector form factors, and no definitive determination of $\Delta s$ was
possible --- this conclusion was confirmed by subsequent reanalyses
of these data~\cite{Garvey:1993cg,Alberico:1998qw}.

The HERMES experiment~\cite{HERMES_deltas}
measured the helicity distribution
of strange quarks, $\Delta s(x)$, using polarized semi-inclusive deep-inelastic
scattering and a leading order ``purity'' analysis, and found 
$\Delta s(x)\approx 0$ in the range $0.03<x<0.3$.  This seems to disagree 
with the analysis of the inclusive deep-inelastic data.  This disagreement 
could be due to a failure of one or more of the assumptions made in the
analysis of the inclusive and/or the semi-inclusive data, or it could be
due to a more exotic physics mechanism such as a ``polarized condensate''
at $x=0$ not observable in deep-inelastic scattering~\cite{BassRMP}.

\section{Combining $ep$ and $\nu p$ Elastic Data}

On account of the apparent discrepancy between the analyses of the two
differnt kinds of deep-inelastic
data, another method is needed to shed light on the strange
quark contribution to the proton spin.  Recently~\cite{Pate:2003rk} it
has become possible to determine the strange vector and axial form
factors of the proton by combining data from elastic 
parity-violating $\vec{e}p$
scattering experiments at Jefferson Lab with the $\nu p$ and $\bar{\nu} p$
elastic scattering data from E734.  The parity-violating $\vec{e}p$ data place
constraints on the strange vector form factors that were not available
for previous analyses of E734 data.

Several experiments have now produced data on forward parity-violating
$\vec{e}p$ elastic scattering
\cite{Aniol:2004hp,Maas:2004ta,Maas:2004dh,
Armstrong:2005hs,HAPPEx_1H_010,HAPPEx2006}.
Of most interest here are measurements that lie in the same $Q^2$
range as the BNL E734 experiment, which are the original HAPPEx
measurement~\cite{Aniol:2004hp} at $Q^2=0.477$~GeV$^2$ and four points
in the recent $G^0$ data~\cite{Armstrong:2005hs}.  These forward
scattering data are most sensitive to $G_E^s$, somewhat less sensitive
to $G_M^s$, and almost completely insensitive to the axial form
factors due to supression by both the weak vector electron charge
$(1-4\sin^2\theta_W)$ and by a kinematic factor that approaches 0 at
forward angles.

The basic technique for combining the $\vec{e}p$, $\nu p$, and $\bar{\nu}p$
data sets has already been
described~\cite{Pate:2003rk} and the details of the present analysis
will be published~\cite{PMP}.  The results are displayed in
Figure~\ref{sff_fig}.  The uncertainties in all three form factors are
dominated by the large uncertainties in the neutrino cross section
data.  Since those data are somewhat insensitive to $G_E^s$ and
$G_M^s$ then the uncertainties in those two form factors are generally
very large.  However the results for the strange axial form factor are of
sufficient precision to give a hint of the $Q^2$-dependence of this
important form factor for the very first time.  There is a strong
indication from this $Q^2$-dependence that $\Delta s < 0$, {\em i.e.}
that the strange quark contribution to the proton spin is negative.
However the data are not of sufficient quality to permit an
extrapolation to $Q^2=0$, so no quantitative evaluation of $\Delta s$
from these data can be made at this time.

\section{Comparison to Model Calculations}

\begin{figure}[t]
  \includegraphics[height=.48\textheight, bb=150 145 495 520]{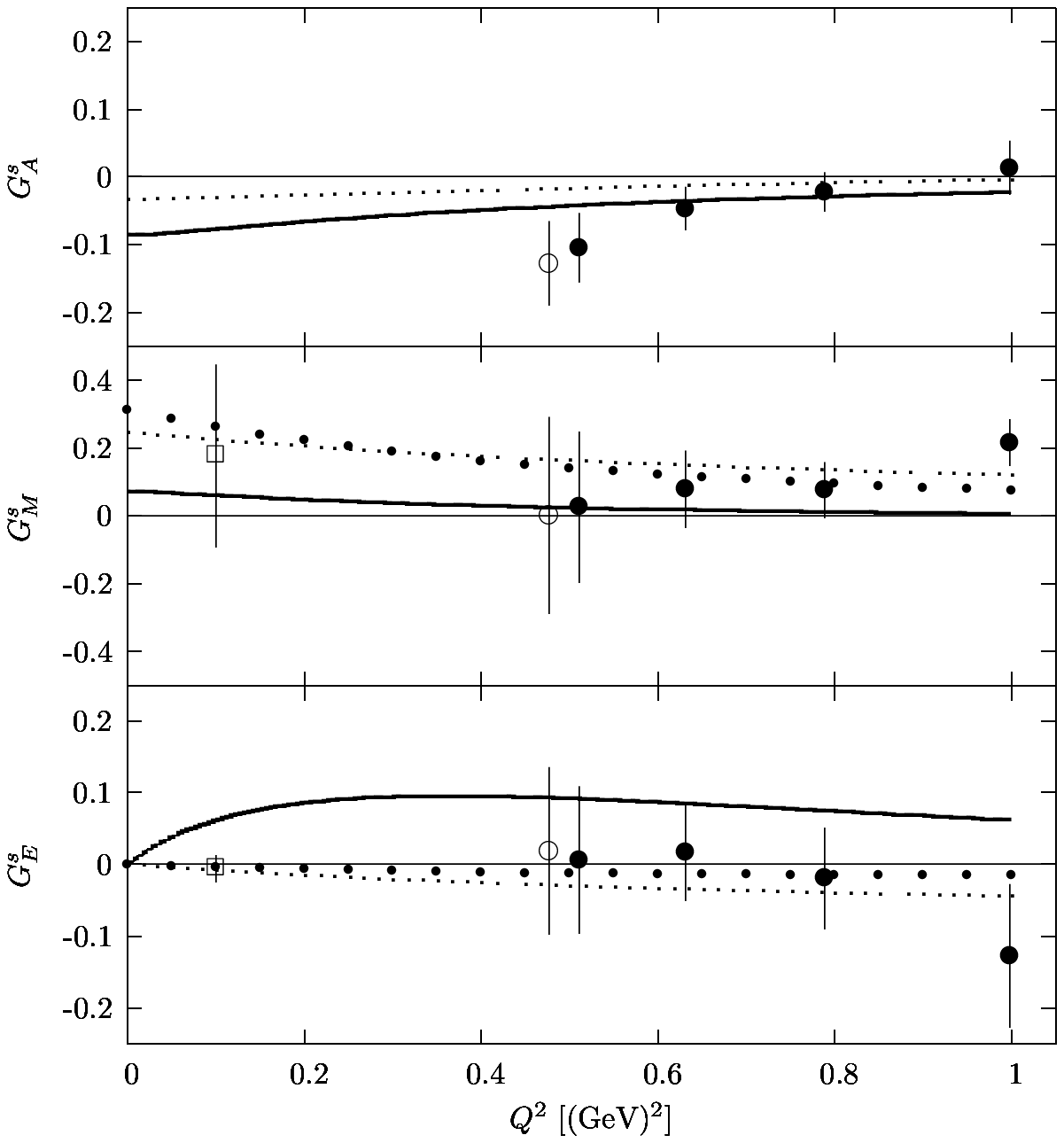}
  \caption{Results of this analysis for the strange vector and 
axial form factors
of the proton.  Open circles are from a combination of 
HAPPEx and E734 data, while the
closed circles are from a combination of $G^0$ and E734 data.  
[Open squares are
from Ref.~\cite{HAPPEx2006} and involve parity-violating $\vec{e}p$ data only.]
The theoretical curves are from 
Ref.~\cite{Silva:2005fa,Goeke:2006gi,Silva:2005qm} (solid line), 
Ref.~\cite{Riska:2005bh} (small-dotted line), and Ref.~\cite{Bijker:2005pe} 
(big-dotted line).  There is not any calculation of $G_A^s$ from 
Ref.~\cite{Bijker:2005pe}.}
\label{sff_fig}
\end{figure}

It is interesting to compare these results with models that can
calculate a $Q^2$-dependence for these form factors.  Silva, Kim,
Urbano and Goeke~\cite{Silva:2005fa,Goeke:2006gi,Silva:2005qm} have
used the chiral quark soliton model ($\chi$QSM) to calculate
$G^s_{E,M,A}(Q^2)$ in the range $0.0<Q^2<1.0$ GeV$^2$.  The $\chi$QSM
has been very successful in reproducing other properties of light
baryons using only a few parameters which are fixed by other data.  In
Figure~\ref{sff_fig} their calculation is shown as the solid line; it
is seen to be in reasonable agreement with the available data,
although the HAPPEx $G^s_E$ point at $Q^2=0.1$ GeV$^2$ disfavors this
calculation. Riska, An, and
Zou~\cite{Zou:2005xy,An:2005cj,Riska:2005bh} have explored the
stangeness content of the proton by writing all possible $uuds\bar{s}$
configurations and considering their contributions to
$G^s_{E,M,A}(Q^2)$.  They find that a unique $uuds\bar{s}$
configuration, with the $s$ quark in a $P$ state and the
$\bar{s}$ in an $S$ state, gives the best fit to the data for
these form factors; see the small-dotted curves in Figure~\ref{sff_fig}.
Bijker~\cite{Bijker:2005pe} uses a two-component model of
the nucleon to calculate $G^s_{E,M}(Q^2)$; the two components are an
intrinsic three-quark structure and a vector-meson ($\rho$, $\omega$,
and $\phi$) cloud; the strange quark content comes from the meson
cloud component.  The values
of $G^s_{E,M}(Q^2)$ are in good agreement with the data, see the
big-dotted line in Figure~\ref{sff_fig}.  In the near future,
the $G^0$ experiment will provided additional data on $G^s_{E,M}(Q^2)$
at 0.23 and 0.63 GeV$^2$ which will help to discriminate between the
$\chi$QSM and the models of Bijker and of Riska et al.

\section{Future Experiments}

To provide a useful determination
of $\Delta s$, better data are needed for both the form factors and the
polarized parton distribution functions.
Two new experiments have been proposed to provide improved neutrino data for
the determination of the strange axial form factor.
FINeSSE~\cite{FINeSSE_FNAL_LOI} proposes to
measure the ratio of the neutral-current
to the charged-current $\nu N$ and $\bar{\nu}N$ processes. A measurement of 
$R_{NC/CC}=\sigma(\nu p\rightarrow\nu p)/\sigma(\nu n\rightarrow\mu^- p)$
and
$\bar{R}_{NC/CC}=\sigma(\bar{\nu}p\rightarrow\bar{\nu}p)/\sigma(\bar{\nu}p\rightarrow\mu^+n)$
combined with the world's data on forward-scattering PV $ep$ data can produce
a dense set of data points for $G_A^s$ in the range $0.25<Q^2<0.75$ GeV$^2$ 
with an uncertainty at each point of about $\pm 0.02$.  
Another experiment with similar physics
goals, called NeuSpin~\cite{NeuSpin}, is being proposed
for the new JPARC facility in Japan.
It is also important to extend the semi-inclusive deep-inelastic data
to smaller $x$ and higher $Q^2$ so that the determination of the
polarized strange quark distribution $\Delta s(x)$ can be improved.  A
measurement of this type is 
envisioned \cite{Stosslein:2000am,Deshpande:2005wd} for the proposed 
electron-ion collider facility.
At the same time, a future electron-ion collider could provide an improved
measurement of the integral $\int_0^1 g_1^p(x)dx$, which can be
combined with a $\nu p$ elastic
scattering measurement of the proton's weak axial charge and 
a recently developed axial charge sum rule~\cite{Bass:2002mv,BassRMP}
to provide an additional determination of $\Delta s$ and perhaps evidence
for a  ``polarized condensate''
at $x=0$~\cite{BassRMP}.
It is only with these improved data sets that we will be able to arrive
at an understanding of the strange quark contribution to the proton spin.

\section{Nuclear Effects in $A(\nu,p)X$}

FINeSSE and NeuSpin will make use of nuclear targets, like
carbon or argon perhaps, in order to bring NC $\nu N \rightarrow \nu N$ 
and CC $\nu N \rightarrow \mu N$ count rates up to a level needed for timely
completion of the experiments.  In 
order to reach the level of precision in $G_A^s$ needed for an extraction
of $\Delta s$ then a good understanding of initial and final state nuclear
effects is needed.  A number of theoretical efforts have been made in the
last few years in this direction.  There is no space here for any detailed
discussion of these calculations but some summary remarks are in order.
Maieron et al.~\cite{Maieron:2003df,Maieron:2004dp} studied quasi-elastic
$\nu_\mu N$ scattering in $^{16}$O and $^{12}$C using a relativistic shell model
and treating final state effects in a distorted wave impulse approximation (DWIA).
Meucci et al.~\cite{Meucci:2004ip,Meucci:2006zr} used a relativistic DWIA model
with a relativistic mean-field model of the nucleus to study the same processes
in $^{12}$C and also studying final state effects.
Martinez et al.~\cite{Martinez:2005xe} studied these processes in $^{12}$C and
$^{56}$Fe using a relativistic description of the nucleus and the scattering 
process; significantly, they argue that nuclear transparencies measured in 
$A(e,e'p)$ experiments can be applied to final state effects in $A(\nu,p)$
processes.  The work of van der Ventel and 
Piekarewicz~\cite{vanderVentel:2003km,vanderVentel:2005ke} centers on the 
calculation of a set of nuclear structure functions
(describing the neutrino-nucleus scattering) via a relativistic
plane wave impulse approximation.
Finally, the work of Leitner et al.~\cite{Leitner:2006ww,Leitner:2006sp} 
uses a fully relativistic formalism including the quasielastic and
resonance scattering channels, and taking into account a variety of nuclear
and final state effects.
The results of several of these studies imply that the ratios $R_{NC/CC}$
and $\bar{R}_{NC/CC}$ to be measured
in FINeSSE are relativly insentive to many effects of the nucleus and final
state interactions.

\begin{acknowledgments}
The author is grateful to S.D. Bass, H.E. Jackson, D.O. Riska, and A.W. Thomas 
for useful discussions.
This work was supported by the US Department of Energy.
\end{acknowledgments}

\bibliography{Cocoyoc_Pate}

\newpage

\end{document}